\documentclass[epj]{svjour}

\renewcommand{\makeheadbox}{\relax} 

\usepackage{graphicx,epstopdf,boxedminipage}
\usepackage{subfig}
\usepackage{caption}
\usepackage{upgreek}
\usepackage{color}
\usepackage{amsmath,amsfonts}
\usepackage{amssymb,bm}

\newcommand{\ket}[1]{\left|{#1}\right>}
\newcommand{\bra}[1]{\left<{#1}\right|}
\newcommand{\I}{\mathrm{i}}
\newcommand{\E}[1]{\mathrm{e}^{\mbox{\footnotesize$#1$}}}
\newcommand{\D}{\mathrm{d}}
\newcommand{\tr}[1]{\mathrm{tr}\!\left\{#1\right\}}
\newcommand{\inner}[2]{\left<{#1}\vphantom{#1}\vphantom{#2}\right|\left.\!{#2}\vphantom{#1}\vphantom{#2}\right>}
\newcommand{\opinner}[3]{\left<{#1}\vphantom{#1}\vphantom{#3}\right|{#2}\left|{#3}\vphantom{#1}\vphantom{#3}\right>}
\newcommand{\StirlingS}[2]{\left[\begin{array}{@{}c@{}}#1\\#2\end{array}\right]}
\newcommand{\StirlingSS}[2]{\left\{\begin{array}{@{}c@{}}#1\\#2\end{array}\right\}}

\newcommand{\BetaBREG}[3]{\mathrm{I}_{#1}\!\left(#2,#3\right)}
\newcommand{\rvec}[1]{\bm{#1}}
\newcommand{\onevec}{\bm{\mathit1}}
\newcommand{\zerovec}{\bm{\mathit0}}
\newcommand{\MEAN}[2]{\mathbb{E}_{#1}\!\left[#2\right]}

\newcommand*\pFqskip{8mu}
\catcode`,\active
\newcommand*\pFq{\begingroup
	\catcode`\,\active
	\def ,{\mskip\pFqskip\relax}%
	\dopFq
}
\catcode`\,12
\def\dopFq#1#2#3#4#5{%
	{}_{#1}F_{#2}\biggl(\begin{array}{@{}c@{}}#3\\#4\end{array};#5\biggr)%
	\endgroup
}

\newcommand*\pFSqskip{8mu}
\catcode`,\active
\newcommand*\pFSq{\begingroup
	\catcode`\,\active
	\def ,{\mskip\pFSqskip\relax}%
	\dopFSq
}
\catcode`\,12
\def\dopFSq#1#2#3#4#5{%
	{}_{#1}F_{#2}\biggl[\begin{array}{@{}c@{}}#3\\#4\end{array};#5\biggr]%
	\endgroup
}

\newcommand*\pFSSqskip{8mu}
\catcode`,\active
\newcommand*\pFSSq{\begingroup
	\catcode`\,\active
	\def ,{\mskip\pFSSqskip\relax}%
	\dopFSSq
}
\catcode`\,12
\def\dopFSSq#1#2#3#4#5{%
	{}_{#1}F_{#2}\biggl\{\begin{array}{@{}c@{}}#3\\#4\end{array};#5\biggr\}%
	\endgroup
}

\allowdisplaybreaks 
\begin{document}

\title{On the theoretical prospects of multiport devices for photon-number-resolving detections}

\author{Yong Siah Teo $^{1}$, Hyunseok Jeong $^{1}$, Jaroslav {\v R}eh{\'a}{\v c}ek $^{2}$, Zden{\v e}k Hradil $^{2}$, Luis L. S{\'a}nchez-Soto $^{3,4}$, and Christine Silberhorn$^5$}

\institute{$^{1}$ \quad Department of Physics and Astronomy, Seoul National University, 08826 Seoul, South Korea\\
	$^{2}$ \quad Department of Optics, Palack\'{y}  University,
	17. listopadu 12, 77146 Olomouc, Czech Republic\\
	$^{3}$ \quad Max-Planck-Institut f\"ur  die Physik des Lichts,
	Staudtstra\ss e 2, 91058 Erlangen, Germany\\
	$^{4}$ \quad Departamento de \'Optica, Facultad de F\'{\i}sica,
	Universidad Complutense, 28040 Madrid, Spain\\
	$^{5}$ \quad Integrated Quantum Optics Group, Applied Physics,
	University of Paderborn, 33098 Paderborn, Germany}

\date{Submitted to arXiv on \today.}

\abstract{Ideal photon-number-resolving detectors form a class of important optical components in quantum optics and quantum information theory. In this article, we theoretically investigate the potential of multiport devices having reconstruction performances approaching that of the Fock-state measurement. By recognizing that all multiport devices are minimally complete, we first provide a general analytical framework to describe the tomographic accuracy (or quality) of these devices. Next, we show that a perfect multiport device with an infinite number of output ports functions as either the Fock-state measurement when photon losses are absent or binomial mixtures of Fock-state measurements when photon losses are present, and derive their respective expressions for the tomographic transfer function. This function is the scaled asymptotic mean squared-error of the reconstructed photon-number distributions uniformly averaged over all distributions in the probability simplex. We then supply more general analytical formulas for the transfer function for finite numbers of output ports in both the absence and presence of photon losses. The effects of photon losses on the photon-number resolving power of both infinite- and finite-size multiport devices are also investigated.
}

\maketitle

\pagestyle{plain}

\onecolumn
\section{Introduction}
Photon-number-resolving (PNR) detection schemes are measurements that play a vital role in quantum information theory. The ability to perform direct photon counting has been shown to fundamentally impact quantum protocols and technologies. These include quantum metrology~\cite{interf1,interf2,interf3,interf4}, quantum key distribution~\cite{QKD1,QKD2}, Bell measurements~\cite{Bell1} and quantum random number generator~\cite{QRNG1,QRNG2}. In practice, such PNR measurements either do not faithfully resolve photon numbers, or do so up to a limited (typically small) number of photons, especially in the emblematic presence of dark counts and photon losses~\cite{PROB1,PROB2}. In recent years, there has been significant progress in the quality and type of photon-counting detectors developed through new-generation quantum engineering techniques~\cite{EXPT1,EXPT2,EXPT3,EXPT4,EXPT5,EXPT6,EXPT7,EXPT8}. 

An alternative class of setups that are widely used to indirectly perform photon counting are the so-called multiport devices~\cite{multiport1,multiport2,multiport3}, which are schematically more sophisticated devices that involve multiple beam splitters and several output ports that lead to ``on--off'' photodetectors for counting the number of split output signal pulses. Such alternative devices are later refashioned using optical-fiber looping~\cite{fiber-loop1,fiber-loop2} or multiplexing~\cite{tmd1,tmd2,tmd3,tmd4} strategies that give exactly the same photon-number-resolving characteristics but with much more efficient and cost-effective architectures. 

In this article, we invoke the machinery of quantum tomography to evaluate the performance of general multiport devices. After providing the general descriptions of multiport devices in Sec.~\ref{sec:multiport_intro} and introducing the concept of informational completeness for such commuting measurements in Sec.~\ref{sec:info_comp}, we establish a general framework in Sec.~\ref{sec:gen_frame} to certify their tomographic performances using an operational \emph{tomographic transfer function} that measures the average asymptotic accuracies of reconstructed photon-number distributions [Eq.~\eqref{eq:TTF2}]. According to this formalism, we first investigate the performances of multiport devices that have infinitely many output ports with and without photon losses in Sec.~\ref{sec:infinite_multiport}. We shall show respectively that these infinitely large devices behave either exactly like a set of Fock-state measurement outcomes or their binomial-noisy mixtures and derive their tomographic transfer functions [Eqs.~\eqref{eq:TTF_fock} and \eqref{eq:TTF_eps_infty}]. We will also demonstrate in Sec.~\ref{subsec:noisy_resolve} that photon losses can severely limit the photon-number resolution of multiport devices and systematically characterize such limitations in terms of informational completeness phase diagrams and the dependence of the maximum photon-loss rate tolerable on the number of photons to be resolved. Finally in Sec.~\ref{sec:finite_multiport}, we shall derive general formulas for the transfer functions for the most general multiport devices with finite output ports [Eqs.~\eqref{eq:TTF_s_perf} and \eqref{eq:TTF_s_imperf}] and evaluate the effects of photon losses on their photon-number resolving power in Sec.~\ref{subsec:noisy_s}.

\section{General physics of multiport devices}
\label{sec:multiport_intro}
A multiport device is a general laboratory equipment that houses an input port for receiving photonic signals and a fixed number (say $s$) of output ports. After undergoing multiple splitting of an input photonic pulse inside the device, each output port would then either idle (symbolically labeled as ``0'') or register a photonic ``click'' (``1'') that originates from the split pulse.
As an example, a three-port device would contain $s=3$ output ports that give a total of $2^3=8$ different detection configurations, which are the ``000'', ``001'', ``010'', ``100'', ``011'', ``101'', ``110'' and ``111'' detection events. For the purpose of photon-number-distribution reconstruction, we may as well consolidate all the ``0-click'', ``1-click'', ``2-click'' and ``3-click'' events respectively and describe this multiport device as a measurement of $M=4$ outcomes.
More generally, an $s$-port device is one that gives $2^s$ detection configurations that may be organized to yield a total of $M=s+1$ measurement events.

Any measurement of quantum sources can be described by a \emph{positive operator-valued measure}~(POVM), a set of probability operators (outcomes) that is given by
\begin{equation}
\Pi_j\geq0 \quad\mathrm{such\,\,that}\quad \sum^{M-1}_{j=0}\Pi_j=1\,.
\end{equation}
A multiport device is no exception, and is therefore mathematically equivalent to a POVM of $M=s+1$ outcomes, where a ``$j$-click'' outcome is some unnormalized mixture $\Pi_j=\sum_n\ket{n}\beta_{jn}\bra{n}$ of Fock states. For sufficiently large number of data sampling events $N$, the data obtained from a measurement of such a POVM give probabilities that are linear combinations of the expectation values $\left<\ket{n}\bra{n}\right>$. The photon-number distribution can subsequently be reconstructed. The amplitudes $\beta_{jn}$ are, in general, complicated functions of \emph{all} the \emph{port efficiencies} $\{\eta_j\}$ $\left(\sum_j\eta_j\leq1\right)$, each of which depends on the physical parameters of the actual device implementation such as beam-splitter ratio, photodetector efficiency, and so on.

In particular, for arbitrary port efficiencies $\sum_j\eta_j\equiv1-\epsilon$, the ``0-click'' outcome $\Pi_0$ possesses amplitudes $\beta_{0,n}=\epsilon^n$ that are independent of any other detail of the multiport specifications. In other words, the probability of a ``0-click'' event for an $n$-photon input signal is the $n$-fold product of the loss probability $\epsilon$, which is consistent with the physical fact that photoabsorption and detector losses are the main mechanisms behind all ``0-click'' events when $n>0$ in the absence of other kinds of experimental imperfections.

If the light source is effectively described by a quantum state $\rho$ in a Hilbert space of dimension $d$, so that the probability of detecting $n>d-1$ photons is practically zero, then all ``$(j>d-1)$-click'' outcomes are correspondingly zero by construction. The outcomes $\Pi_j$ are hence represented by $d\times d$ positive matrices that sum to the identity matrix. We can define the \emph{measurement matrix} that concisely and uniquely determine the multiport POVM. To do this, we first emphasize that in this effective Hilbert space, the conditional photon-number probabilities $\rho_n=\opinner{n}{\rho}{n}=\left<\ket{n}\bra{n}\right>$ are properly normalized $(\tr{\rho}=1)$, so that the total number of independent parameters to be estimated is $d-1$. From Born's rule, we may express the multiport probabilities in terms of $d-1$ independent state parameters inasmuch as
\begin{eqnarray}
p_j=\tr{\rho\Pi_j}&=&\,\sum^{d-1}_{n=0}\beta_{jn}\rho_n\nonumber\\
&=&\,\sum^{d-2}_{n=0}\beta_{jn}\rho_n+\beta_{j\,d-1}\left(1-\sum^{d-2}_{n=0}\rho_{n}\right)\nonumber\\
&=&\,\sum^{d-2}_{n=0}\left(\beta_{jn}-\beta_{j\,d-1}\right)\rho_n+\beta_{j\,d-1}\,.
\end{eqnarray}
Following the reasonings in quantum-state tomography \cite{TTF,ysteo-book}, we may define the measurement matrix
\begin{equation}
\bm{C}=\sum^{d-1}_{j=0}\sum^{d-2}_{n=0}\rvec{e}_j\rvec{e}_n\left(\beta_{jn}-\beta_{j\,d-1}\right)\,,
\end{equation}
with the help of the standard computational basis $\rvec{e}_l\bm{\cdot}\rvec{e}_{l'}=\delta_{l,l'}$, to be the $d\times(d-1)$ rectangular matrix that fully characterizes the multiport POVM for the $d-1$ independent $\rho_n$ parameters. It is clear that this matrix has a zero eigenvalue corresponding to the eigenvector $\onevec_{\!d}$ that is represented as a $d$-dimensional column of ones---$\onevec_{\!d}\bm{\cdot}\bm{C}=\zerovec_{\!d-1}^{\textsc{t}}$.

We shall look into an interesting special case where the port efficiencies are all equal to a constant ($\eta_j=\eta$), so that the POVM amplitudes can be shown to take the simple form~\cite{multiport1,fiber-loop1,tmd1}
\begin{equation}
\beta_{jn}=(-1)^j\binom{s}{j}\sum^j_{k=0}\binom{j}{k}(-1)^k\left[1-\eta(s-k)\right]^n\,.
\label{eq:amp_unif_eff}
\end{equation}
The self-consistent consequence that
\begin{eqnarray}
\sum^{s}_{j=0}\beta_{jn}&=&\,\sum^{s}_{j=0}(-1)^j\binom{s}{j}\sum^j_{k=0}\binom{j}{k}(-1)^k\left[1-\eta(s-k)\right]^n\nonumber\\
&=&\,\sum^{s}_{k=0}(-1)^k\left[1-\eta(s-k)\right]^n\sum^{s}_{j=k}(-1)^j\binom{s}{j}\binom{j}{k}\nonumber\\
&=&\,\sum^{s}_{k=0}(-1)^k\left[1-\eta(s-k)\right]^n\underbrace{(-1)^k\binom{s}{k}\sum^{s-k}_{j'=0}(-1)^{j'}\binom{s-k}{j'}}_{\qquad\quad\,\,\,\,\,\,\,\displaystyle{=(-1)^s\,\delta_{s,k}}}\nonumber\\
&=&\,1
\end{eqnarray}
can be verified straightforwardly. This type of multiport device is commonly used in practice. We mention in passing that the outcomes $\Pi_j$ may be equivalently expressed as the normal-ordered form
\begin{equation}
\Pi_j=\binom{s}{j}\,\bm{:}\!\left(\E{-\eta a^\dagger a}\right)^{s-j}\left(1-\E{-\eta a^\dagger a}\right)^{j}\bm{:}
\label{eq:op_normord}
\end{equation}
from which Eq.~\eqref{eq:amp_unif_eff} is quickly obtained through the application of the formula
\begin{equation}
\bm{:}F(a^\dagger a)\bm{:}\,\,=\left.F\!\left(\frac{\D}{\D x}\right)x^{a^\dagger a}\right|_{x=1}
\end{equation}
for any operator function $F(a^\dagger a)$ of the number operator $a^\dagger a$. 

Dark counts may be incorporated in a simplistic way by introducing the parameter $\nu>0$ that defines the average dark-count rate as the transformation $\eta a^\dagger a\rightarrow \eta a^\dagger a+\nu$ to Eq.~\eqref{eq:op_normord}. Physically, this transformation increases the partially-depleted number operator $\eta a^\dagger a$ due to losses by an additional $\nu$ photons on average. In what follows, dark-count rates are assumed to be negligible in the feasible bandwidth of the photodetectors.

\section{Informational completeness of photon-number distribution measurements}
\label{sec:info_comp}

To analyze photon-number-distribution reconstruction with multiport devices, we shall review the tools that are employed in understanding quantum measurements in this context. We recall that an informationally complete (IC) measurement is one that \emph{uniquely} characterizes a particular set of physically relevant parameters describing a given quantum source of interest. In quantum-state tomography, such a measurement unambiguously reconstructs the quantum state $\rho$ for the source. For our purpose, the set of parameters constitutes the photon-number distribution $\{\rho_n\}$ of a quantum light source, which are the diagonal entries of $\rho$ in the Fock basis as mentioned in Sec.~\ref{sec:multiport_intro}. With respect to the $\rho_n$s, a POVM is IC when it contains at least $d$ outcomes with a degree of linear independence of $d$.

The entire machinery for IC quantum-state tomography can be translated for photon-number distribution tomography. The concept of the \emph{operator ket} is particularly helpful here for notational simplification. For any $d$-dimensional operator $O$ in the Fock basis, its operator ket $\ket{O}$ is defined as the $d$-dimensional column vector of its diagonal entries. The photon-number distribution of $\rho$ that is of interest to us is thus summarized by its operator ket $\ket{\rho}$ such that $\tr{\rho}=\inner{1}{\rho}=1$. With this, we can define the \emph{frame operator}
\begin{equation}
\mathcal{F}=\sum^{M-1}_{j=0}\frac{\ket{\Pi_j}\bra{\Pi_j}}{\tr{\Pi_j}}
\label{eq:frame}
\end{equation}
for any POVM $\{\Pi_j\}$ comprising $M$ commuting Fock-state mixtures. Hence, an equivalent definition for an IC POVM is the operator invertibility of $\mathcal{F}$. In addition, using the operator-ket notation, the degree of linear independence of the POVM can be checked by inspecting the eigenvalues of the standard Gram matrix
\begin{equation}
\bm{G}=\sum^{M-1}_{j=0}\sum^{M-1}_{k=0}\rvec{e}_j\rvec{e}_k\inner{\Pi_j}{\Pi_k}=\rvec{V}\rvec{V}^\dagger\,,\quad
\rvec{V}=\left(\begin{array}{@{}c@{}}
\bra{\Pi_1}\\
\vdots\\
\bra{\Pi_M}
\end{array}\right)\,,
\label{eq:Gram}
\end{equation}
for vectorial objects.

For multiport devices, Eq.~\eqref{eq:frame} is applicable for $M=s+1$. Consequently, it is necessary for the corresponding multiport POVM to have $s\geq d-1$ output ports for it to be IC in a $d$-dimensional Hilbert space. Moreover, there exists another important feature for these devices. As discussed in Sec.~\ref{sec:multiport_intro}, that the probability of detecting more photons than the number available in the input signal is zero implies that \emph{any} multiport POVM is necessarily \emph{minimally complete} when it is IC on the $d$-dimensional Hilbert space. This means that for such minimal POVMs, there are effectively only $M=d$ nonzero outcomes (each having amplitudes that depend on $s$) and we can \emph{uniquely} express the photon-number distribution as
\begin{equation}
\ket{\rho}=\sum^{d-1}_{j=0}\ket{\Theta_j}p_j
\end{equation}
with the help of the $d$ \emph{canonical dual operators}
\begin{equation}
\ket{\Theta_j}=\mathcal{F}^{-1}\frac{\ket{\Pi_j}}{\tr{\Pi_j}}\,.
\label{eq:dual_op}
\end{equation}
It can be shown that
\begin{equation}
\inner{\Pi_j}{\Theta_k}=\delta_{j,k}
\label{eq:pom_dual_orth}
\end{equation}
for any minimal POVM. For this, we use the general property
\begin{equation}
\rvec{W}^\dagger\rvec{V}=1=\rvec{V}^\dagger\rvec{W}\,,\quad\rvec{W}=\left(\begin{array}{@{}c@{}}
\bra{\Theta_1}\\
\vdots\\
\bra{\Theta_M}
\end{array}\right)\,,
\label{eq:dual_op_prop}
\end{equation}
for any set of (canonical) dual operators, so that sandwiching the left equation in \eqref{eq:dual_op_prop} with $\rvec{V}$ from the left and $\rvec{V}^\dagger$ from the right gives
\begin{equation}
\rvec{V}\rvec{W}^\dagger\rvec{G}=\rvec{G}=\rvec{G}\rvec{W}\rvec{V}^\dagger\,.
\end{equation}
Next, we realize that for any minimal POVM, $\bm{G}$ is always invertible and we have $\rvec{V}\rvec{W}^\dagger=1=\rvec{W}\rvec{V}^\dagger$.

\section{General framework for the reconstruction accuracy of multiport devices}
\label{sec:gen_frame}

\subsection{Mean squared-error and its Cram{\'e}r--Rao bound}

We shall take the mean squared-error (MSE) $\mathcal{D}_\textsc{mse}$ as the measure of the reconstruction accuracy of the photon-number distribution $\ket{\rho}$. For a given estimator $\ket{\widehat{\rho}}$ of $\ket{\rho}$, since only $\rho_n\big|^{d-2}_{n=0}$ are \emph{independent}, this measure is defined as
\begin{equation}
\mathcal{D}_\textsc{mse}=\MEAN{\mathrm{data}}{\left(\ket{\widehat{\rho}}-\ket{\rho}\right)^2}\bigg|_{\mathrm{sup}}\,,
\end{equation}
\noindent
where the average is taken over all plausible data. The label ``sup'' means that the inner product is evaluated in the $(d-1)$-dimensional support of the linearly independent parameters of $\ket{\rho}$. The parameter space of $\ket{\rho}$ is the entire $d$-dimensional probability simplex, since one can always find a quantum state $\rho$ that gives any particular $\ket{\rho}$ (a statistical mixture of Fock states weighted with the $\rho_n$s, for instance). The boundary of this space is therefore the edges of this simplex.

When $\ket{\rho}$ is off the boundary ($\rho_n\neq0$), which is the real experimental situation, it is well-known that the scaled MSE with $N$ is bounded from below by the Cram{\'e}r--Rao bound (CRB) per sampling event,
\begin{equation}
N\mathcal{D}_\textsc{mse}\geq\tr{F(\rho)^{-1}}\,,
\end{equation}
\noindent
where $F(\rho)$ is the $(d-1)$-dimensional Fisher information operator (defined per sampling event) for a given $\ket{\rho}$ and
POVM. In particular, the unbiased \emph{maximum-likelihood} (ML) estimator saturates this bound asymptotically in the limit of large $N$. Boundary $\ket{\rho}$s may be included in the picture by taking appropriate limits. The CRB directly evaluates the reconstruction accuracy of $\ket{\rho}$ where the constraint $\tr{\rho}=\inner{1}{\rho}=1$ is obeyed, and supplies the limit of photon-number reconstruction for any $\ket{\rho}$.

For any minimal POVM, the MSE has a simple compact form for single-shot experiments that yield multinomial data statistics, just as for any multiport device. First, we can define the \emph{linear estimator} of $\ket{\rho}$, in terms of the canonical dual operators and the measured multiport relative frequencies $\nu_j$, as
\begin{equation}
\ket{\widehat{\rho}}=\sum^{d-1}_{j=0}\ket{\Theta_j}\nu_j\,,
\label{eq:lin_est}
\end{equation}
where $\inner{\Pi_j}{\rho}=\nu_j$ for any minimal POVM. The fact that $\MEAN{\mathrm{data}}{\ket{\widehat{\rho}}}=\ket{\rho}$ is evident. Second, we recall that this linear estimator is in fact the ML estimator whenever $\ket{\rho}>\zerovec$ for sufficiently large $N$, so that the linear estimator in Eq.~\eqref{eq:lin_est} saturates the CRB. So, using the identity
\begin{equation}
\MEAN{\mathrm{data}}{\nu_j\nu_k}=\frac{1}{N}[\delta_{j,k}p_j+(N-1)p_jp_k]
\end{equation}
for multinomial distributions, we have
\begin{eqnarray}
\mathcal{D}_\textsc{mse}&=&\,\MEAN{\mathrm{data}}{\inner{\widehat{\rho}}{\widehat{\rho}}}-\inner{\rho}{\rho}\Big|_{\mathrm{sup}}\nonumber\\
&=&\,\left.\sum^{d-1}_{j=0}\sum^{d-1}_{k=0}\inner{\Theta_j}{\Theta_k}\left(\MEAN{\mathrm{data}}{\nu_j\nu_k}-p_jp_k\right)\right|_{\mathrm{sup}}\nonumber\\
&=&\,\left.\frac{1}{N}\left(\sum^{d-1}_{j=0}\inner{\Theta_j}{\Theta_j}p_j-\inner{\rho}{\rho}\right)\right|_{\mathrm{sup}}\,.
\end{eqnarray}
On the other hand for multinomial data statistics, it is known that the Fisher operator takes the form
\begin{eqnarray}
F(\ket{\rho})&=&\,\left.\sum^{d-1}_{l=0}\left(\ket{\Pi_l}-\ket{1}\beta_{l\,d-1}\right)\frac{1}{p_l}\left(\bra{\Pi_l}-\beta_{l\,d-1}\bra{1}\right)\right|_{\mathrm{sup}}\nonumber\\
&=&\,\bm{C}^\textsc{t}\bm{P}^{-1}\bm{C}\,,
\label{eq:Fisher_mnorm}
\end{eqnarray}
where $\bm{P}_j=p_j$. In view of this, we arrive at the identity
\begin{eqnarray}
\tr{F(\ket{\rho})^{-1}}&=&\,\tr{\left(\bm{C}^\textsc{t}\bm{P}^{-1}\bm{C}\right)^{-1}}\nonumber\\
&=&\,\left.\left(\sum^{d-1}_{j=0}\inner{\Theta_j}{\Theta_j}p_j-\inner{\rho}{\rho}\right)\right|_{\mathrm{sup}}
\label{eq:CRB_multiport}
\end{eqnarray}
\noindent
for any minimal POVM with respect to the photon-number distribution.

\subsection{A measure of tomographic performance}

The CRB in Eq.~\eqref{eq:CRB_multiport} is a function of $\ket{\rho}$. To obtain an operational performance certifier, one may choose to average over $\ket{\rho}$, which can be carried out in many different ways. We shall follow a similar direction reported in Ref.~\cite{TTF} and perform an average over all distributions over the probability simplex. The resulting average CRB
\begin{equation}
\mathrm{TTF}=\MEAN{\ket{\rho}}{\tr{F(\ket{\rho})^{-1}}}
\end{equation}
\noindent
is the \emph{tomographic transfer function} (TTF) for photon-number distributions, which generalizes previous analytical scopes, such as those in~\cite{PROB2} and \cite{multiport3}, that focus on the class of Poissonian distributions to other more exotic yet classically allowed probability distributions $\{p_j\}$ in the $(d-1)$-dimensional simplex. Following through the calculations, using the simplex identities (see Appendix~\ref{app:simplex})
\begin{equation}
\MEAN{\ket{\rho}}{p_j}=\frac{1}{d}\quad\text{and}\quad\MEAN{\ket{\rho}}{p_j^2}=\frac{2}{d(d+1)}\,,
\end{equation}
we have
\begin{equation}
\MEAN{\ket{\rho}}{\inner{\rho}{\rho}}\Big|_{\mathrm{sup}}=\sum^{d-2}_{n=0}\MEAN{\ket{\rho}}{\rho_n^2}=\frac{2(d-1)}{d(d+1)}\,.
\end{equation}
Finally\footnote{It turns out that Eq.~\eqref{eq:TTF1} may also be obtained from an average of $F(\ket{\rho})$ uniformly (under the Haar measure) over all pure states $\rho$.}, after a reference to Eq.~\eqref{eq:dual_op},
\begin{equation}
\mathrm{TTF}_\mathrm{multiport}=\MEAN{\ket{\rho}}{\tr{F(\rho)^{-1}}}=\frac{1}{d}\tr{\mathcal{F}^{-1}}\Big|_{\mathrm{sup}}-\frac{2(d-1)}{d(d+1)}\,.
\label{eq:TTF1}
\end{equation}
One can proceed to express the first term on the rightmost side of Eq.~\eqref{eq:TTF1} by recognizing that
\begin{equation}
\tr{\mathcal{F}^{-1}}\Big|_{\mathrm{sup}}=\tr{\mathcal{F}^{-1}}-\opinner{d-1}{\mathcal{F}^{-1}}{d-1}\,,
\end{equation}
and that the Fock state
\begin{equation}
\ket{d-1}=\frac{\ket{\Pi_{d-1}}}{\tr{\Pi_{d-1}}}
\end{equation}
for any multiport device since in the absence of dark counts, the ``$j$-click'' event occurs when there are $j$ photons or more. Then the orthonormality property in Eq.~\eqref{eq:pom_dual_orth} dictates that
\begin{equation}
\opinner{d-1}{\mathcal{F}^{-1}}{d-1}=\frac{1}{\tr{\Pi_{d-1}}^2}\opinner{\Pi_{d-1}}{\mathcal{F}^{-1}}{\Pi_{d-1}}=\frac{1}{\tr{\Pi_{d-1}}}\,,
\end{equation}
which brings us to the slightly more explicit expression
\begin{equation}
\mathrm{TTF}_\mathrm{multiport}(s,\{\eta_j\})=\frac{1}{d}\tr{\mathcal{F}^{-1}}-\frac{1}{d\,\tr{\Pi_{d-1}}}-\frac{2(d-1)}{d(d+1)}\,.
\end{equation}
The result in Eq.~\eqref{eq:TTF2} assigns a number to the average performance of a multiport device of arbitrary number of output ports $s$, port efficiencies $\{\eta_j\}$ and loss probability $\epsilon$ based on statistical estimation theory.

For any multiport POVM of amplitudes $\beta_{jn}$, by defining $\bm{B}_{s,\epsilon}$ to be the square matrix of these amplitudes $(\bm{B}_{s,\epsilon})_{jn}=\beta_{jn}$, the operator kets $\ket{\Pi_j}$ and the Fock kets $\ket{n}$ are then related by the simple linear system
\begin{equation}
\rvec{V}=\bm{B}_{s,\epsilon}\rvec{v}\,,\quad\rvec{v}=\left(\begin{array}{@{}c@{}}
\bra{0}\\
\vdots\\
\bra{d-1}
\end{array}\right)\,,
\end{equation}
where the column $\rvec{V}$ of operator bras is as defined in Eq.~\eqref{eq:Gram}. The Fock bras can then be expressed in terms of the operator bras $\ket{\Pi_j}$ as $\rvec{v}=\bm{B}_{s,\epsilon}^{-1}\rvec{V}$. This compact form proves useful when evaluating the operator trace of the inverted frame operator $\mathcal{F}^{-1}$:
\begin{equation}
\tr{\mathcal{F}^{-1}}=\tr{\mathcal{F}^{-1}\rvec{v}^\dagger\rvec{v}}=\tr{\mathcal{F}^{-1}\rvec{V}^\dagger{\bm{B}_{s,\epsilon}^\textsc{t}}^{-1}\bm{B}_{s,\epsilon}^{-1}\rvec{V}}\,.
\end{equation}
At this stage, we emphasize the distinction between the operators (such as $\mathcal{F}$ and $\ket{\Pi_j}\bra{\Pi_j}$) and the columns of (operator)~kets (such as $\rvec{v}$ and $\rvec{V}$) to avoid confusion regarding the role of the operator trace $\tr{\,\cdot\,}$. With that, using the basic fact
\begin{equation}
\opinner{\Pi_j}{\mathcal{F}^{-1}}{\Pi_k}=\tr{\Pi_j}\delta_{j,k}
\label{eq:min_POVM_opinner}
\end{equation}
for any minimal POVM, the answer
\begin{equation}
\tr{\mathcal{F}^{-1}}=\sum^{d-1}_{j=0}\tr{\Pi_j}\left(\bm{B}_{s,\epsilon}\bm{B}_{s,\epsilon}^\textsc{t}\right)^{-1}_{jj}
\end{equation}
is immediate and
\begin{equation}
\mathrm{TTF}_\mathrm{multiport}(s,\{\eta_j\})=\frac{1}{d}\left[\sum^{d-1}_{j=0}\tr{\Pi_j}\left(\bm{B}_{s,\epsilon}\bm{B}_{s,\epsilon}^\textsc{t}\right)^{-1}_{jj}-\frac{1}{\tr{\Pi_{d-1}}}-\frac{2(d-1)}{d+1}\right]\!.
\label{eq:TTF2}
\end{equation}

\section{Multiport device of equal port efficiencies and $s\rightarrow\infty$ output ports}
\label{sec:infinite_multiport}

To gain some physical insights from the structure of multiport devices, we begin with a systematic study of the special case where $\eta_j=\eta$. With this, Eq.~\eqref{eq:amp_unif_eff} immediately applies. Upon an introduction of the simple relation
\begin{equation}
\left[1-\eta(s-k)\right]^n=\left.\left(\frac{\partial}{\partial t}\right)^n\E{t\left[1-\eta(s-k)\right]}\right|_{t=0}\,,
\end{equation}
subsequent analysis may be facilitated after rewriting the POVM amplitudes as
\begin{equation}
\beta_{jn}=\left.\binom{s}{j}\left(\frac{\partial}{\partial t}\right)^n\left[\E{t(1-\eta s)}\left(\E{t\eta}-1\right)^j\right]\right|_{t=0}\,.
\label{eq:amp_unif_eff_2}
\end{equation}
This formula, which is valid for any $s$ and $\eta$, shall serve as a good starting point for deriving our main results.

\subsection{Perfect multiport devices without losses}

If the loss probability is zero ($\epsilon=0$), the port efficiencies are then all equal to $\eta=1/s$. It follows from Eq.~\eqref{eq:amp_unif_eff_2}, that
\begin{equation}
\beta_{jn}=\left.\binom{s}{j}\left(\frac{\partial}{\partial t}\right)^n\left(\E{\frac{t}{s}}-1\right)^j\right|_{t=0}=\frac{j!}{s^n}\binom{s}{j}\StirlingSS{n}{j}
\label{eq:amp_unif_eff_perf}
\end{equation}
after an invocation of the moment-generating formula
\begin{equation}
\left.\frac{1}{a^nj!}\left(\frac{\partial}{\partial t}\right)^n\left(\E{at}-1\right)^j\right|_{t=0}=\StirlingSS{n}{j}
\label{eq:Rodrigues_StirlingS2}
\end{equation}
for the Stirling number of the second kind $\StirlingSS{n}{j}$. The combinatorial sum rule
\begin{equation}
\sum^n_{j=0}\frac{s!}{(s-j)!}\StirlingSS{n}{j}=s^n
\end{equation}
guarantees the proper normalization of $\beta_{jn}$ as it should.

For infinitely many ports ($s\rightarrow\infty$), the ratio $s!/(s-j)!\rightarrow s^j$ and the amplitudes
\begin{equation}
\beta_{jn}\rightarrow\left.\frac{1}{s^{n-j}}\StirlingSS{n}{j}\right|_{s\rightarrow\infty}=\delta_{j,n}
\end{equation}
\noindent
become those of the Fock states. Put differently, as the multiport device grows in size, its functionality approaches that of the pure Fock-state measurement---indirect photon counting approaches direct photon counting in the large-$s$ limit. As the matrix $\bm{B}\equiv\bm{B}_{s\rightarrow\infty,\epsilon=0}$ is simply the $d\times d$ identity matrix, the TTF takes the value
\begin{equation}
\mathrm{TTF}_\mathrm{multiport}\left(s\rightarrow\infty,\left\{\eta_j=\frac{1}{s}\right\}\right)=\frac{(d-1)^2}{d(d+1)}\,.
\label{eq:TTF_fock}
\end{equation}
\noindent
As $d$ increases, the TTF approaches unity. It can be shown that the performance $\mathrm{TTF}_\mathrm{multiport}\left(s,\left\{\eta_j\right\}\right)$ of any arbitrary lossless multiport device is bounded from below by this Fock-state limit (see Appendix~\ref{app:fock_opt}).

\subsection{Imperfect multiport devices with losses}

When photon losses are present $[\epsilon>0,\eta=(1-\epsilon)/s]$, Eq.~\eqref{eq:amp_unif_eff_2} gives
\begin{equation}
\beta_{jn}=\left.\binom{s}{j}\left(\frac{\partial}{\partial t}\right)^n\left\{\E{t\epsilon}\left[\E{\frac{t}{s}(1-\epsilon)}-1\right]^j\right\}\right|_{t=0}\,.
\end{equation}
In the limit $s\rightarrow\infty$, the approximation $\E{y}\approx1+y$ for small $y$ and $s!/(s-j)!\rightarrow s^j$ render
\begin{eqnarray}
\beta_{jn}&\rightarrow&\,\left.\frac{\left(1-\epsilon\right)^j}{j!}\left(\frac{\partial}{\partial t}\right)^n\left(t^j\E{t\epsilon}\right)\right|_{t=0}=\frac{\left(1-\epsilon\right)^j}{j!}\sum^\infty_{l=0}\frac{\epsilon^{\,l}}{l!}\!\!\!\!\underbrace{\left.\left(\frac{\partial}{\partial t}\right)^nt^{j+l}\right|_{t=0}}_{\displaystyle{\qquad\quad\,=n!\,\delta_{l,n-j}}}\nonumber\\
&=&\,\binom{n}{j}(1-\epsilon)^j\epsilon^{n-j}=(\bm{B}_\epsilon)_{jn}\,.
\end{eqnarray}
These amplitudes correspond to those of a POVM comprising \emph{binomial mixtures of Fock-state outcomes}\footnote{The expression for $\bm{B}_\epsilon$ was defined earlier in \cite{tmd2} as a separate consequence of multiport photon losses, the argument of which is independent of taking the limit $s\rightarrow\infty$.}
\begin{equation}
\Pi_j=\sum^{d-1}_{m=j}\ket{m}\binom{m}{j}(1-\epsilon)^j\,\epsilon^{m-j}\bra{m}\,,
\end{equation}
which tends to the set of Fock states in the limits $\epsilon\rightarrow0$ and $d\rightarrow\infty$. Thus for large multiport devices, the probabilities are primarily influenced by the number of detection and absorption events.

To calculate the TTF, we need the inverse of $\bm{B}_{\epsilon}\equiv\bm{B}_{s\rightarrow\infty,\epsilon}$, which can be deduced to be
\begin{equation}
\bm{B}_{\epsilon}^{-1}=\sum^{d-1}_{j=0}\sum^{d-1}_{n=0}\rvec{e}_{j}\rvec{e}_{n}\binom{n}{j}(1-\epsilon)^{-n}(-\epsilon)^{n-j}
\label{eq:Binv_eps}
\end{equation}
by a reverse engineering of the binomial theorem. One can effortlessly verify the following obvious necessary property $\bm{B}_{\epsilon}^{-1}\bm{B}_{\epsilon}=\bm{1}$.
The remaining task is to simply calculate the matrix elements of $\left(\bm{B}_{\epsilon}\bm{B}_{\epsilon}^\textsc{t}\right)^{-1}$:
\begin{eqnarray}
\left(\bm{B}_{\epsilon}\bm{B}_{\epsilon}^\textsc{t}\right)^{-1}_{jj'} &=&\, \sum^{d-1}_{n=0}\binom{j}{n}(1-\epsilon)^{-j}(-\epsilon)^{j-n}\,\binom{j'}{n}(1-\epsilon)^{-j'}(-\epsilon)^{j'-n}\nonumber\\
&=&\,\left(-\frac{\epsilon}{1-\epsilon}\right)^{j'+j}\sum^{d-1}_{n=0}\binom{j}{n}\binom{j'}{n}\frac{1}{\epsilon^{2n}}\nonumber\\
&=&\,\left(-\frac{\epsilon}{1-\epsilon}\right)^{j'+j}\pFq{2}{1}{-j,-j'}{1}{\frac{1}{\epsilon^2}}\,,
\label{eq:BBTinv_eps_infty}
\end{eqnarray}
where we have considered a definition
\begin{equation}
\sum^{j_<}_{n=0}\binom{j}{n}\binom{j'}{n}y^n=\pFq{2}{1}{-j,-j'}{1}{y}\,,\quad j_<=\min\{j,j'\}\,,
\label{eq:hypergeom_def}
\end{equation}
for the special case of the Gaussian hypergeometric function $\pFq{2}{1}{a_1,a_2}{b_1}{y}$.

\begin{figure}[t]
	\centering
	\includegraphics[width=0.7\columnwidth]{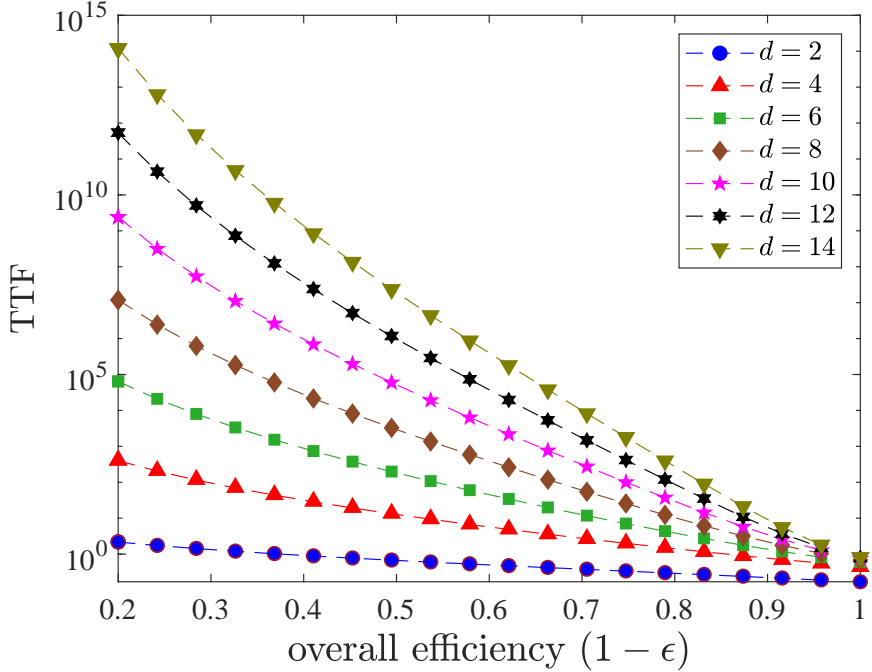}
	\caption{\label{fig:TTF_th_sim}Numerical~(colored markers) and theoretical~(colored dashed curves) values of TTF (logarithmically scaled) for infinitely large multiport devices of various $\epsilon$ and Hilbert-space dimensions $d$. A total of 1000 random pure states were used to evaluate each numerical plot point. The convergence to the optimal TTF in Eq.~\eqref{eq:TTF_fock} at $\epsilon=1$ is as expected. It therefore comes as no surprise that losses monotonically lowers reconstruction accuracy.}
\end{figure}

Furthermore, in terms of the regularized incomplete beta function $\BetaBREG{z}{a}{b}$, the operator traces for this multiport POVM
\begin{equation}
\tr{\Pi_j}=\frac{1}{1-\epsilon}\left[1-\BetaBREG{\epsilon}{d-j}{j+1}\right]\,.
\label{eq:trPi}
\end{equation}
Notably, we have $\tr{\Pi_{d-1}}=(1-\epsilon)^{d-1}$, which can be obtained either by
\begin{equation}
\pFq{2}{1}{1,d+1}{2}{y}=\frac{1}{dy}\left[\frac{1}{(1-y)^d}-1\right]
\end{equation}
or the simple physical reasoning that the registration of $j$ clicks must at least originate from the presence of $j$ photons ($j\leq n$). A substitution of this final piece of information as well as Eq.~\eqref{eq:BBTinv_eps_infty} into Eq.~\eqref{eq:TTF2} leads to
\begin{eqnarray}
&&\,\mathrm{TTF}_\mathrm{multiport}\left(s\rightarrow\infty,\left\{\eta_j=\frac{1-\epsilon}{s}\right\}\right)\nonumber\\
&=&\,\frac{1}{d}\Bigg[\sum^{d-1}_{j=0}\tr{\Pi_j}\left(\frac{\epsilon}{1-\epsilon}\right)^{2j}\pFq{2}{1}{-j,-j}{1}{\frac{1}{\epsilon^2}}-\frac{1}{(1-\epsilon)^{d-1}}-\frac{2(d-1)}{d+1}\Bigg]\,.
\label{eq:TTF_eps_infty}
\end{eqnarray}
\noindent
It is clear that $\epsilon=0$ brings us back to the optimal result stated in Eq.~\eqref{eq:TTF_fock} by noting that
\begin{equation}
\left.\epsilon^{2j}\,\pFq{2}{1}{-j,-j}{1}{\frac{1}{\epsilon^2}}\right|_{\epsilon=0}=1
\end{equation}
that arises from the definition in Eq.~\eqref{eq:hypergeom_def}.

Figure~\ref{fig:TTF_th_sim} demonstrates the fit between the theoretically predicted TTF values with Eq.~\eqref{eq:TTF_eps_infty} and the numerically calculated ones after performing Monte-Carlo averaging of the inverse of the Fisher operator $F(\rho)^{-1}$ [see Eq.~\eqref{eq:Fisher_mnorm}] over the Haar measure of pure states. To this average numerically, it is sufficient to generate a sufficiently large number of random pure states $\left\{\rho_j=\mathcal{A}^\dagger_j\mathcal{A}_j/\tr{\mathcal{A}^\dagger_j\mathcal{A}_j}\right\}$ parametrized by the random complex auxiliary rank-one operators $\mathcal{A}_j$ that follow the standard Gaussian distribution and use them to compute the average of $F(\rho)^{-1}$.

\begin{figure}[t]
	\centering
	\includegraphics[width=1\columnwidth]{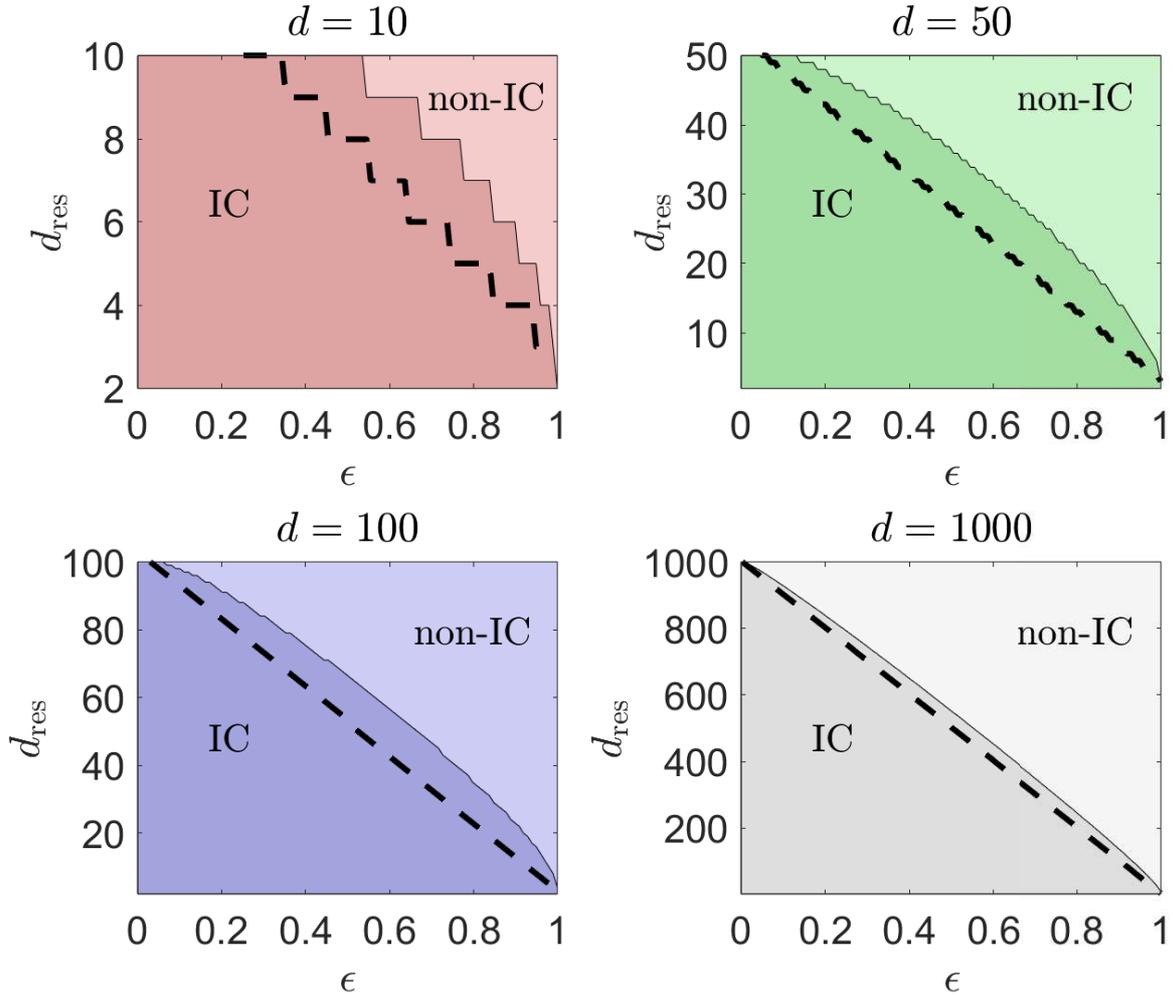}
	\caption{\label{fig:phase_diag}Informational completeness phase diagrams for various $d$ in the $d_\mathrm{res}$-$\epsilon$ plane with $\mu_\mathrm{thres}=10^{-3}$. Subspaces of dimensions below the boundary are resolvable, and hence render the multiport device of $s\rightarrow\infty$ and $\eta_j=(1-\epsilon)/s$ IC. Those of dimensions above the boundary are unresolvable with such a multiport device. The thick dashed curves represent the analytically calculated boundaries using the approximation in \eqref{eq:analytic_approx}, which provide conservative underestimates for the maximum $d_\mathrm{res}$ compared to the numerically computed boundaries. Clearly, the range of $\epsilon$ for which the entire $d$-dimensional Hilbert space is completely resolvable reduces as $d$ increases.}
\end{figure}

\subsection{Noisy photon-number resolution of multiport devices with $s\rightarrow\infty$ and $\epsilon>0$}
\label{subsec:noisy_resolve}

In the hypothetical situation where the photon-loss rate $\epsilon=0$, the multiport device is capable of resolving photon numbers in an optical signal described by $\rho$ of any arbitrary dimension $d$\footnote{Recall that $d-1$ is then the maximum number of photons in the signal}. We say that the $d$-dimensional Hilbert space is resolvable. Therefore any subspace of dimension $d_\mathrm{res}\leq d$ is by definition also resolvable. In real experiments however, a nonzero photon-loss rate directly limits the number of photons resolvable. The key relation that governs this restriction for $s\rightarrow\infty$ is Eq.~\eqref{eq:trPi}. For a fixed $d$, $\tr{\Pi_j}$ (or $\Pi_j$) becomes essentially zero above certain threshold $j=j_\mathrm{thres}$. This threshold value defines the dimension of the maximally resolvable subspace---$d_\mathrm{res}\leq j_\mathrm{thres+1}$. 

More specifically, we may define $j_\mathrm{thres}$ as the largest integer for which
\begin{equation}
1-\BetaBREG{\epsilon}{d-j_\mathrm{thres}}{j_\mathrm{thres}+1}>\mu_\mathrm{thres}\approx 0\,,
\end{equation}
where $\mu_\mathrm{thres}$ is a very small positive number close to zero. While this equation has no general analytical solution for finite $d$, we note that for large $d$, $\BetaBREG{\epsilon}{d-j}{j+1}\approx\frac{1}{2}+\frac{1}{2}\tanh(j-d(1-\epsilon))$ is a remarkably good approximation. We may then use this to derive the simplified and approximate photon-number resolvability restriction
\begin{equation}
d_\mathrm{res}\leq d(1-\epsilon)+\tanh^{-1}(1-2\mu_\mathrm{thres})\,.
\label{eq:analytic_approx}
\end{equation}

\begin{figure}[t]
	\centering
	\includegraphics[width=0.6\columnwidth]{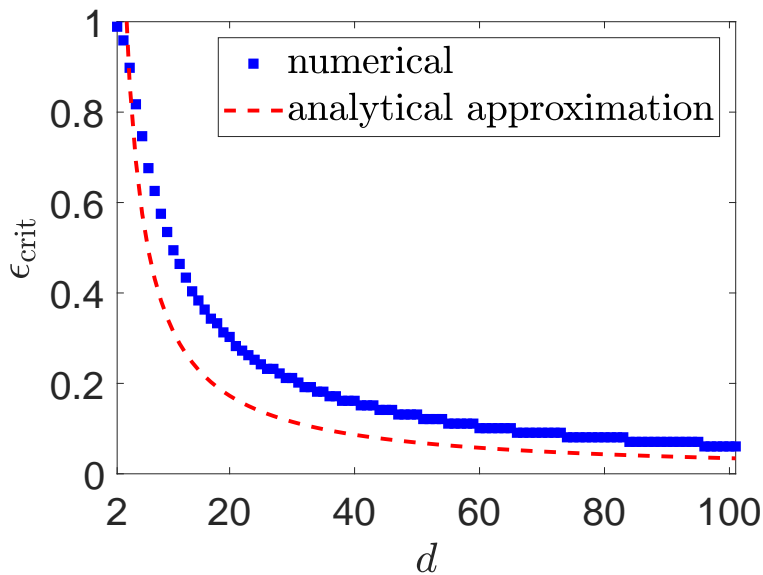}
	\caption{\label{fig:eps_crit}A plot of the critical $\epsilon$ value ($\epsilon_\mathrm{crit}$) against $d$ that shows the maximum amount of photon losses a multiport device can tolerate before losing its informational completeness property. Here, $\mu_\mathrm{thres}=10^{-3}$. The simple $\epsilon_\mathrm{crit}\sim 1/d$ behavior serves as a back-of-the-envelope solution for designing such devices.}
\end{figure}

This observation impacts how we should perform asymptotic TTF analyses for multiport devices in the large $d$-limit. Unlike the ideal case where one simply takes $d\rightarrow\infty$ with Eq.~\eqref{eq:TTF_fock} to arrive at the finite value 1, this naive limit results in the divergence of $\mathrm{TTF}_\mathrm{multiport}$ for any finite $\epsilon$. A careful thought reveals that indeed, for the tomography of photon-number distributions for dimension $d$ to be IC, we require the \emph{necessary condition} that $\epsilon$ be no greater than some critical value beyond which the inequality $1-\BetaBREG{\epsilon}{1}{d}>\mu_\mathrm{thres}$ becomes in valid. This condition may be approximately written as
\begin{equation}
\epsilon\leq \tanh^{-1}(1-2\mu_\mathrm{thres})/d
\end{equation}
for sufficiently large $d$ following Eq.~\eqref{eq:analytic_approx}. Figures \ref{fig:phase_diag} and \ref{fig:eps_crit} show the important plots that characterize the informational completeness of any given (infinitely large) multiport device of nonzero photon-loss rate $\epsilon$.

\section{Multiport device of equal port efficiencies and $s$ output ports}
\label{sec:finite_multiport}

\subsection{Perfect multiport devices without losses}

The consideration of a finite-size multiport device with $s$ output ports more closely resembles the real physical situation in the laboratory in which the resources that go into its implementation are limited. Even in this case, one can still easily compute the TTF for the $\epsilon=0$ case where losses are absent in the device. Starting with the POVM amplitudes $\beta_{jn}$ in Eq.~\eqref{eq:amp_unif_eff_perf}, we find that the inverse of their corresponding $\bm{B}_{s}\equiv\bm{B}_{s,\epsilon=0}$ amplitude matrix is simply given by
\begin{equation}
\bm{B}_{s}^{-1}=\sum^{d-1}_{j=0}\sum^{d-1}_{n=0}\rvec{e}_j\rvec{e}_n\frac{(s-n)!}{s!}s^j(-1)^{n-j}\StirlingS{n}{j}
\label{eq:Binv_s}
\end{equation}
and we owe this simple inversion formula to the existence of the (unsigned) Stirling number of the first kind $\StirlingS{n}{j}$ that is orthogonal to the Stirling number of the second kind $\StirlingSS{n}{j}$ in the sense that
\begin{equation}
\sum^k_{n=j}(-1)^{n-k}\StirlingSS{n}{j}\StirlingS{k}{n}=\sum^k_{n=j}(-1)^{n-j}\StirlingSS{k}{n}\StirlingS{n}{j}=\delta_{j,k}\,.
\end{equation}
This means that
\begin{eqnarray}
\left(\bm{B}_{s}\bm{B}_s^\textsc{t}\right)^{-1}_{jj'}&=&\,(-1)^{j+j'}\frac{(s-j)!\,(s-j')!}{s!^2}\sum^{d-1}_{n=0}s^{2n}\StirlingS{j}{n}\StirlingS{j'}{n}\nonumber\\
&=&\,(-1)^{j+j'}\frac{(s-j)!\,(s-j')!}{s!^2}\,\pFSq{2}{1}{-j,-j'}{1}{s^2}\,,
\end{eqnarray}
where we have defined the Stirling--Gaussian hypergeometric function of the first kind
\begin{equation}
\sum^{j_<}_{n=0}\StirlingS{j}{n}\StirlingS{j'}{n}y^n=\pFSq{2}{1}{-j,-j'}{1}{y}
\label{eq:stirl1_hypergeom_def}
\end{equation}
that is of analogous form to the usual Gaussian hypergeometric function in Eq.~\eqref{eq:hypergeom_def}. Accordingly, the Stirling--Gaussian hypergeometric function $\pFSSq{2}{1}{-j,-j'}{1}{y}$ of the second kind would then simply involve the Stirling numbers of the second kind.

\begin{figure}[t]
	\centering
	\includegraphics[width=0.7\columnwidth]{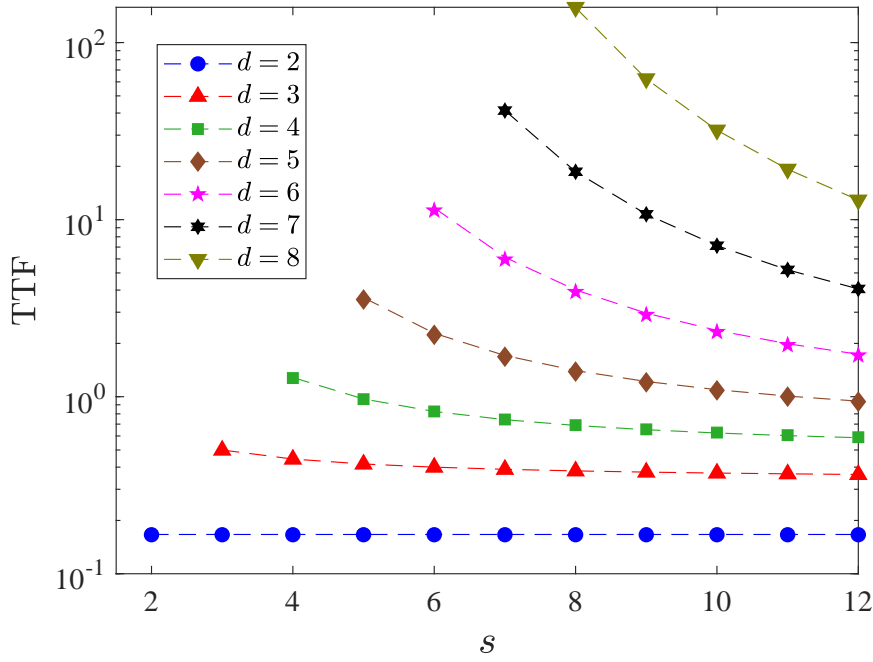}
	\caption{\label{fig:TTF_s_th_sim}Numerical~(colored markers) and theoretical~(colored dashed curves) values of the TTF (logarithmically scaled) for finite-size multiport devices of various $s$ values and Hilbert-space dimensions $d$. A total of 2000 random pure states were used to evaluate each numerical plot point. The $s\geq d$ regime illustrates the TTF for IC multiport POVMs only, which is the regime an observer would be interested in for the purpose of photon-number-distribution tomography.}
\end{figure}

The resulting performance certifier
\begin{eqnarray}
&&\,\mathrm{TTF}_\mathrm{multiport}\left(s,\left\{\eta_j=\frac{1}{s}\right\}\right)\nonumber\\
&=&\,\frac{1}{d}\Bigg\{\sum^{d-1}_{j=0}\frac{(s-j)!}{s!}\,\pFSq{2}{1}{-j,-j}{1}{s^2}\sum^{d-1}_{n'=j}\frac{1}{s^{n'}}\StirlingSS{n'}{j}-\frac{s^{d-1}(s-d+1)!}{s!}-\frac{2(d-1)}{d+1}\Bigg\}
\label{eq:TTF_s_perf}
\end{eqnarray}
\noindent
allows us to evaluate the reconstruction accuracy for a finite-size multiport device of equal port efficiencies. As a verification of the validity of Eq.~\eqref{eq:TTF_s_perf}, we compare it with numerically computed TTF for a sufficiently large set of random pure states distributed to the Haar measure (see~Fig.~\ref{fig:TTF_s_th_sim}).
Specifically, we note that for $d=2$, the TTF is a constant value of $1/6$, which tells us that for effective single-photon sources a two-port device functions exactly like a Fock-state measurement. This can be easily understood in hindsight by realizing that the only POVM outcomes that matter in this subspace are the vacuum and $n=1$ Fock states in the absence of losses. All other $s-1$ outcomes are not measured.

\subsection{Imperfect multiport devices with losses}

\begin{figure}[t]
	\centering
	\includegraphics[width=0.7\columnwidth]{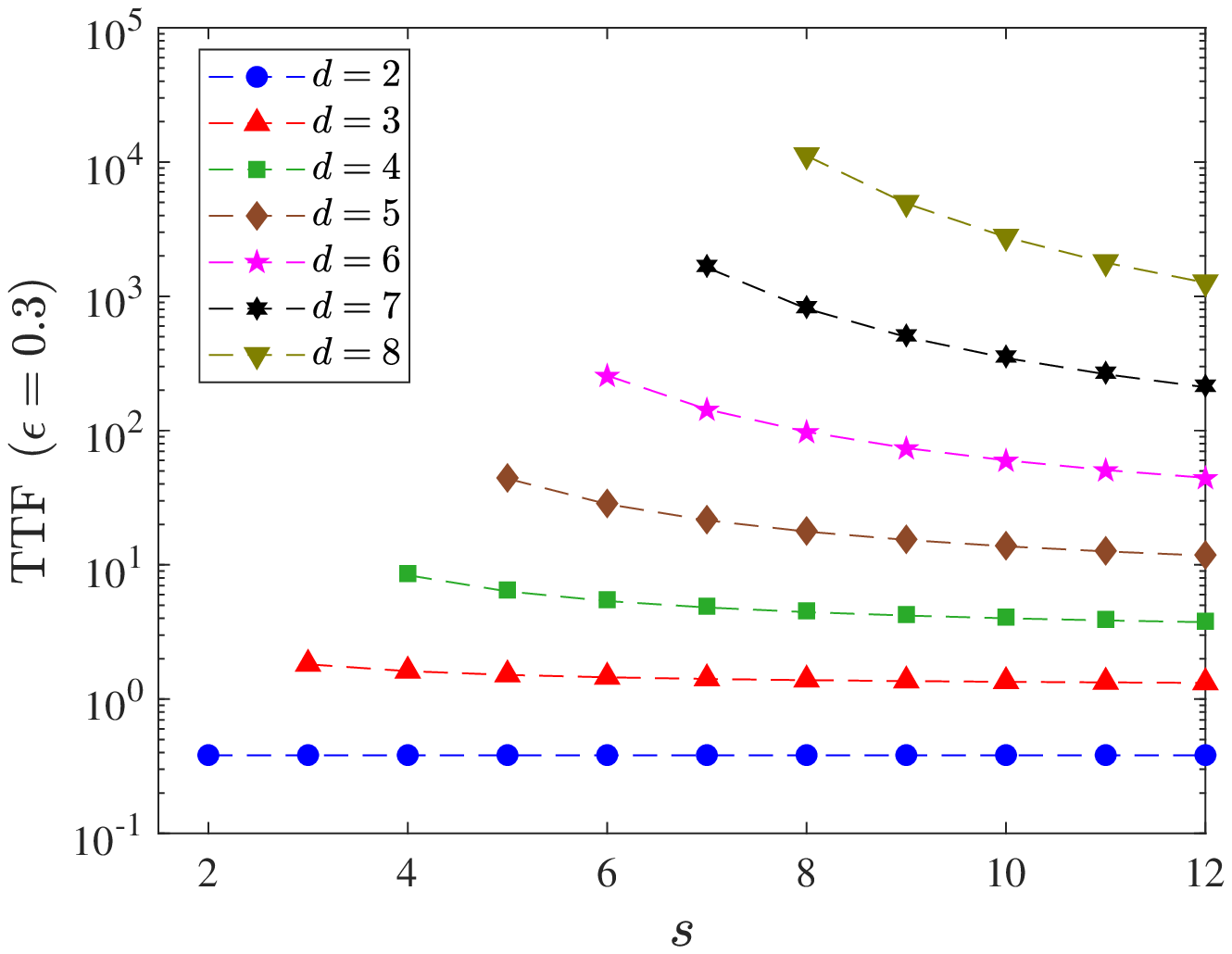}
	\caption{\label{fig:TTF_s_eps_th_sim}Numerical~(colored markers) and theoretical~(colored dashed curves) values of the TTF (logarithmically scaled) for finite-size multiport devices of various $s$ values, a fixed $\epsilon=0.3$, and Hilbert-space dimensions $d$. A total of 2000 random pure states were used to evaluate each numerical plot point. As in the case of $\epsilon=0$, the TTF for $d=2$ takes a constant value of 0.3809 for this particular $\epsilon$ value. The worsening of the tomographic performance with a finite loss probability is clearly manifested as an overall increase in the TTF values.}
\end{figure}

The rather specialized physical and mathematical structures of multiport devices permit us to obtain an analytical expression for the TTF even in the most general case where $s<\infty$ and $\epsilon>0$. The corresponding $\bm{B}_{s,\epsilon}$ for such multiport POVMs can again be inverted by the observation that $\bm{B}_{s,\epsilon}=\bm{B}_s\bm{B}_\epsilon$. In other words, a finite-size photoabsorptive multiport device is a device convolution of a perfect finite-size multiport device and photoabsorption losses. This is because
\begin{eqnarray}
\left(\bm{B}_s\bm{B}_\epsilon\right)_{jn}&=&\,\sum^{d-1}_{n'=0}\frac{1}{s^{n'}}\frac{s!}{(s-j)!}\StirlingSS{n'}{j}\binom{n}{n'}(1-\epsilon)^{n'}\epsilon^{n-n'}\nonumber\\
&=&\,\epsilon^{n}\binom{s}{j}\sum^{n}_{n'=0}\left.\binom{n}{n'}\left(\frac{1}{\epsilon}\frac{\partial}{\partial t}\right)^{n'}\left[\E{\frac{t(1-\epsilon)}{s}}-1\right]^j\right|_{t=0}\nonumber\\
&=&\,(-1)^j\epsilon^{n}\binom{s}{j}\sum^{j}_{k=0}\binom{j}{k}(-1)^k\!\!\!\!\!\!\!\!\!\underbrace{\sum^{n}_{n'=0}\binom{n}{n'}\left(\frac{1-\epsilon}{\epsilon s}k\right)^{n'}}_{\displaystyle{\qquad\qquad\qquad\quad=\left(1+\frac{1-\epsilon}{\epsilon s}k\right)^n}}\nonumber\\
&=&\,(-1)^j\binom{s}{j}\sum^{j}_{k=0}\binom{j}{k}(-1)^k\left(\epsilon+\frac{1-\epsilon}{s}k\right)^n\nonumber\\
&=&\,(-1)^j\binom{s}{j}\sum^{j}_{k=0}\binom{j}{k}(-1)^k\left[1-\eta(s-k)\right]^n\nonumber\\
&=&\,\left(\bm{B}_{s,\epsilon}\right)_{jn}\,.
\end{eqnarray}
This decomposition implies that $\bm{B}_{s,\epsilon}^{-1}=\bm{B}_\epsilon^{-1}\bm{B}_s^{-1}$, so that utilizing the results from Eqs.~\eqref{eq:Binv_eps} and \eqref{eq:Binv_s} for the two respective components, we can summarize the expressions for the performance measure:

\begin{eqnarray}
\mathrm{TTF}_\mathrm{multiport}\left(s,\left\{\eta_j=\frac{1-\epsilon}{s}\right\}\right)&=&\,\frac{1}{d}\Bigg[\sum^{d-1}_{j=0}\tr{\Pi_j}\left(\bm{W}^\textsc{t}_{s,\epsilon}\bm{W}_{s,\epsilon}\right)_{jj}-\frac{1}{\tr{\Pi_{d-1}}}-\frac{2(d-1)}{d+1}\Bigg]\,,\nonumber\\
\bm{W}_{s,\epsilon\,\,jn}&=&\,(-1)^{n-j}\epsilon^{-j}\frac{(s-n)!}{s!}\sum^n_{l=j}\left(\frac{\epsilon s}{1-\epsilon}\right)^l\binom{l}{j}\StirlingS{n}{l}\,.
\label{eq:TTF_s_imperf}
\end{eqnarray}

\noindent
Once more, we notice the constant TTF for $d=2$ with a value of $(1+2\epsilon)/(6-6\epsilon)$ due to the $s$-independent multiport POVM consisting of the outcomes
\begin{eqnarray}
\Pi_0 &=&\,\ket{0}\bra{0}+\ket{1}\epsilon\bra{1}\,,\nonumber\\
\Pi_1 &=&\,\ket{1}(1-\epsilon)\bra{1}\,.
\end{eqnarray}
Figure~\ref{fig:TTF_s_eps_th_sim} gives the comparison between theory and numerical computations for a sample $\epsilon$.

\subsection{Noisy photon-number resolution of multiport devices with $s<\infty$ and $\epsilon>0$}
\label{subsec:noisy_s}

As with the case of $s\rightarrow\infty$ in Sec.~\ref{subsec:noisy_resolve}, a nonzero photon-loss rate $\epsilon$ for a finite-size multiport device also reduces the number of photons that can be resolved. Therefore, $\epsilon$ should again be smaller than some critical value in order for the multiport device to characterize the complete $d$-dimensional photon-number distribution. This critical value may be computed, for every given value of $s$, according to the constraint $\sum^{d-1}_{n=0}(\bm{B}_{s,\epsilon})_{d-1,n}>\mu_\mathrm{thres}\approx0$ that is to be satisfied by the largest value of $\epsilon$.

\begin{figure}[t]
	\centering
	\includegraphics[width=0.6\columnwidth]{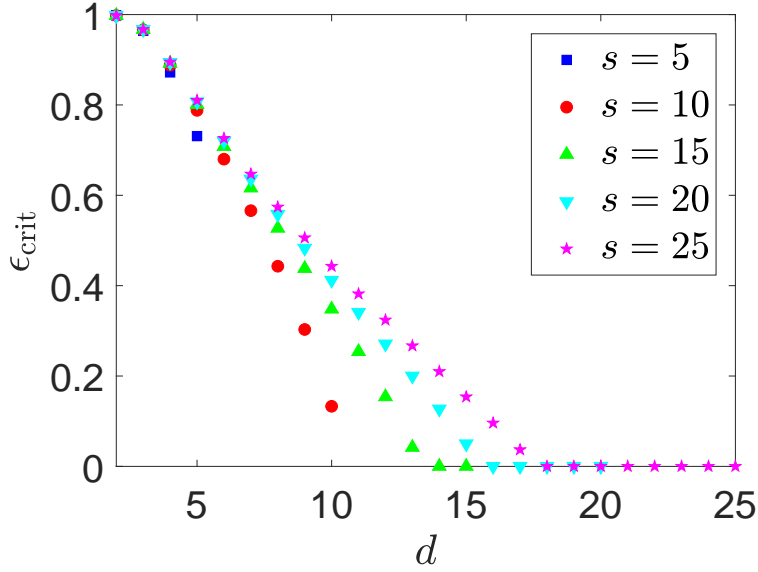}
	\caption{\label{fig:eps_crit_s}A plot of the critical $\epsilon$ value ($\epsilon_\mathrm{crit}$) against $d$ for various number of outputs $s$ of the multiport device. The threshold $\mu_\mathrm{thres}=10^{-3}$ is chosen. For a given $d$-dimensional Hilbert subspace, increasing $s$ also raises $\epsilon_\mathrm{crit}$, although for reasonable values of $s$ such an increase is not dramatic even when $d\ll s$.}
\end{figure}

In general, the critical value of $\epsilon$ has no easy analytical form. It is however numerically efficient to plot graphs of the critical values with respect to $d$ for any physically reasonable $s$. Figure~\ref{fig:eps_crit_s} shows some sample plots.

\section{Discussion}

We present a short series of studies related to the performance of multiport devices on photon-number distribution tomography. The central measure of performance is the quantum tomographic transfer function---the uniform average of the inverse Fisher information over all photon-number distributions in the probability simplex. 

The mathematical framework for calculating the transfer function introduced in this article allows us to conclude that a sufficiently-large multiport devices of equal transmissivity for each output port function either like a Fock-state measurement or binomial mixtures of Fock-state measurements respectively in the absence and presence of photon losses. These are followed by analytical treatments for finite-size multiport devices. In the presence of photon losses, we have studied and mapped out conditions concerning the photon-number resolving power of noisy multiport devices of both infinite and finite sizes. We show that devices of high photon losses possess weak photon-number resolving power and increasing the number of output ports may help only to a certain limited extent. The optimization of photodetectors and other optical components, especially for the purpose of curbing photon losses, is therefore crucial for building realistic multiport devices for indirect photon counting.

\section*{Acknowledgments}
We acknowledge financial support from the BK21 Plus Program (Grant No. 21A20131111123) funded by the Ministry of Education (MOE, Korea) and National Research Foundation of Korea (NRF), the NRF grant funded by the Korea government (MSIP) (Grant No. 2010-0018295), the Spanish MINECO (Grant No. FIS2015-67963-P), the Grant Agency of the Czech Republic (Grant No. 18-04291S), and the IGA Project of the Palack{\'y} University (Grant No. IGA PrF 2019-007).

\appendix
\section{Optimality of the Fock-state measurement for noiseless multiport devices}
\label{app:fock_opt}

We expect the commuting Fock-state measurement to be the optimal noiseless measurement for photon-number-distribution reconstruction. This expectation can be confirmed by showing that the value of the TTF in Eq.~\eqref{eq:TTF_fock} is indeed the optimal limit for all multiport devices. To this end, we exploit the inequalities
\begin{eqnarray}
\qquad\qquad\,\,\,\,\tr{AB}&\leq&\,\tr{A}\tr{B}\quad\mathrm{for\,\,}A\geq0\mathrm{\,\,and\,\,}B\geq0\,,\label{eq:ineq1}\\
\qquad\tr{A}\tr{A^{-1}}&\geq&\,\mathrm{dim}\{A\}^2\quad\mathrm{for\,\,any\,\,invertible\,\,}A\,,
\label{eq:ineq2}
\end{eqnarray}
and remind ourselves that the TTF expression in Eq.~\eqref{eq:TTF2} holds for any $s$, $\{\eta_j\}$ and $\epsilon$. As the operator trace
\begin{equation}
\tr{\mathcal{F}}=\sum^{d-1}_{j=0}\frac{\tr{\Pi_j^2}}{\tr{\Pi_j}}\leq\sum^{d-1}_{j=0}\tr{\Pi_j}=\tr{1}=d
\end{equation}
of the general frame operator is bounded from above according to \eqref{eq:ineq1}, the inequality in \eqref{eq:ineq2} implies that
\begin{equation}
\tr{\mathcal{F}^{-1}}\geq d\,.
\end{equation}
Together with the obvious fact that the probability of detecting all available photons from the input signal never exceeds one ($\tr{\Pi_{d-1}}\leq1$), we have the general inequality
\begin{equation}
\mathrm{TTF}_\mathrm{multiport}(s,\{\eta_j\})\geq\mathrm{TTF}_\mathrm{multiport}\left(s\rightarrow\infty,\left\{\eta_j=\frac{1}{s}\right\}\right)
\end{equation}
\noindent
to confirm that the Fock-state measurement condition $\{s\rightarrow\infty,\left\{\eta_j=1/s\right\}\}$ is indeed optimal.

\section{Averages over the probability simplex}
\label{app:simplex}

We shall give simple derivations of the identities in \eqref{eq:TTF1}. The general $m$-moment integral of interest in our context takes the form
\begin{equation}
I_m=\int^1_0\,\D p_0\cdots\int^1_0\,\D p_{d-1}\,\delta\!\left(1-\sum^{d-1}_{l=0}p_l\right)\,p_j^m\,,
\end{equation}
in which the simplex constraint $\sum^{d-1}_{l=0}p_l=1$ is obeyed. Using the integral representation 
\begin{equation}
\delta(x)=\int\dfrac{\D\,k}{2\pi}\,\E{\I k x}
\end{equation}
for the delta function,
\begin{align}
	I_m=&\,\int\dfrac{\D\,k}{2\pi}\,\E{\I k}\int^1_0\,\D p_0\cdots\int^1_0\,\D p_{d-1}\,\E{-\I k(p_0+\cdots+p_{d-1})}\,p_j^m\nonumber\\
	=&\,\int\dfrac{\D\,k}{2\pi}\,\E{\I k}\,\dfrac{\big(1-\E{-\I k}\big)^{d-1}}{(\I k)^{d-1}}\int^1_0\,\D p_j\,\E{-\I kp_j}\,p_j^m\nonumber\\
	=&\,\int\dfrac{\D\,k}{2\pi}\,\E{\I k}\,\dfrac{\big(1-\E{-\I k}\big)^{d-1}}{(\I k)^{d-1}}\left(\I\dfrac{\D}{\D k}\right)^m\left[\dfrac{1}{\I k}\left(1-\E{-\I k}\right)\right]\,,
\end{align}
where if $y\equiv\I k$,
\begin{align}
	\left(\I\dfrac{\D}{\D k}\right)^m\left[\dfrac{1}{\I k}\left(1-\E{-\I k}\right)\right]=&\,(-1)^m\left(\dfrac{\D}{\D y}\right)^m\left[y^{-1}(1-\E{-y})\right]\nonumber\\
	=&\,m!\,y^{-m-1}(1-\E{-y})-\sum^m_{n=1}\dfrac{m!}{n!}\,y^{-m+n-1}\,\E{-y}\,,
\end{align}
so that
\begin{equation}
I_m=m!\int\dfrac{\D\,k}{2\pi}\,\E{\I k}\,\dfrac{\big(1-\E{-\I k}\big)^{d}}{(\I k)^{m+d}}-\sum^m_{n=1}\dfrac{m!}{n!}\,\int\dfrac{\D\,k}{2\pi}\,\dfrac{\big(1-\E{-\I k}\big)^{d-1}}{(\I k)^{m-n+d}}\,.
\label{eq:Im_working}
\end{equation}

The first term can be evaluated using the identity
\begin{equation}
\dfrac{1}{y^{m+1}}=\dfrac{1}{m!}\int^\infty_0\,\D t\,t^m\,\E{-y t}\,.
\end{equation}
This gives
\begin{align}
	m!\int\dfrac{\D\,k}{2\pi}\,\E{\I k}\,\dfrac{\big(1-\E{-\I k}\big)^{d}}{(\I k)^{m+d}}=&\,\dfrac{m!}{(m+d-1)!}\sum^{d}_{n=0}\binom{d}{n}(-1)^n\int^\infty_0\,\D t\,t^{m+d-1}\underbrace{\int\dfrac{\D\,k}{2\pi}\,\E{\I k(1-n-t)}}_{\displaystyle=\delta(1-n-t)}\nonumber\\
	=&\,\dfrac{m!}{(m+d-1)!}\sum^{d}_{n=0}\binom{d}{n}(-1)^n\,\delta_{n,0}\nonumber\\
	=&\,\dfrac{m!}{(m+d-1)!}\,.
	\label{eq:res_first}
\end{align}
Next, it is possible to argue that the second term of \eqref{eq:Im_working} is zero, since upon repeating the same exercise, we arrive at $\delta(n+t)$ instead of $\delta(1-n-t)$ as in the first line of \eqref{eq:res_first}.

Therefore, the $m$th moment of $p_j$ over the probability simplex is
\begin{equation}
\MEAN{\ket{\rho}}{p_j^m}=\dfrac{I_m}{I_0}=\binom{m+d-1}{m}^{-1}\,.
\end{equation}
In the regime of $d\gg m$, we then have $\MEAN{\ket{\rho}}{p_j^m}\approx\dfrac{\sqrt{2\pi m}}{(\E{}d/m)^m}=O\left(\dfrac{1}{d^m}\right)$.



\end{document}